\documentclass[%
 aip,
 amsmath,amssymb,
reprint,%
]{revtex4-1}

\usepackage{graphicx}% Include figure files
\usepackage{dcolumn}% Align table columns on decimal point
\usepackage{bm}% bold math
\usepackage[utf8]{inputenc}
\usepackage[T1]{fontenc}
\usepackage{mathptmx}

\begin{document}

\preprint{AIP/123-QED}

\title[]{Thermal neutron cross sections of amino acids from average contributions of functional groups}
% Force line breaks with \\

\author{Giovanni Romanelli}
 \affiliation{ISIS Neutron and Muon Source, UKRI-STFC, Rutherford Appleton Laboratory, Harwell Campus, Didcot, Oxfordshire OX11 0QX, United Kingdom}
\author{Dalila Onorati}%
\email[Corresponding Author: ]{dalila.onorati@uniroma2.it}
\affiliation{Universit{\`a} degli Studi di Roma ``Tor Vergata'', Dipartimento di Fisica and NAST Centre, Via della Ricerca Scientifica 1, Roma 00133, Italy}

\author{Pierfrancesco Ulpiani}
\affiliation{Universit{\`a} degli Studi di Roma ``Tor Vergata'', Dipartimento di Scienze e Tecnologie Chimiche, Via della Ricerca Scientifica 1, Roma 00133, Italy}

\author{Stephanie Cancelli}
\affiliation{Universit{\`a} di Milano-Bicocca, Piazza della Scienza 3, Milano, Italy }

\author{Enrico Perelli-Cippo}
\affiliation{Universit{\`a} di Milano-Bicocca, Piazza della Scienza 3, Milano, Italy }

\author{Jos{\'e} Ignacio M{\'a}rquez Dami{\'a}n}
\affiliation{European Spallation Source ERIC, P.O. Box 176, 22100 Lund, Sweden}

\author{Silvia C. Capelli}
\affiliation{ISIS Neutron and Muon Source, UKRI-STFC, Rutherford Appleton Laboratory, Harwell Campus, Didcot, Oxfordshire OX11 0QX, United Kingdom}

\author{Gabriele Croci}
\affiliation{Universit{\`a} di Milano-Bicocca, Piazza della Scienza 3, Milano, Italy }
\affiliation{Istituto per la Scienza e Tecnologia dei Plasmi, CNR, via Cozzi 53, 20125 Milano, Italy}

\author{Andrea Muraro}
\affiliation{Istituto per la Scienza e Tecnologia dei Plasmi, CNR, via Cozzi 53, 20125 Milano, Italy}

\author{Marco Tardocchi}
\affiliation{Istituto per la Scienza e Tecnologia dei Plasmi, CNR, via Cozzi 53, 20125 Milano, Italy}

\author{Giuseppe Gorini}
\affiliation{Universit{\`a} di Milano-Bicocca, Piazza della Scienza 3, Milano, Italy }

\author{Carla Andreani}
\affiliation{Universit{\`a} degli Studi di Roma ``Tor Vergata'', Dipartimento di Fisica and NAST Centre, Via della Ricerca Scientifica 1, Roma 00133, Italy}
\affiliation{CNR-ISM, Area della Ricerca di Roma Tor Vergata, Via del Fosso del Cavaliere 100, 00133 Roma, Italy}

\author{Roberto Senesi}
\affiliation{Universit{\`a} degli Studi di Roma ``Tor Vergata'', Dipartimento di Fisica and NAST Centre, Via della Ricerca Scientifica 1, Roma 00133, Italy}
\affiliation{CNR-IPCF, Sezione di Messina, Viale Ferdinando Stagno d'Alcontres 37, Messina, 98158, Italy}

\date{\today}% It is always \today, today,
             %  but any date may be explicitly specified

\begin{abstract}
The experimental thermal neutron cross sections of the twenty proteinogenic amino acids have been measured over the incident-neutron energy range spanning from 1 meV to 10 keV and data have been interpreted using the multi-phonon expansion based on first-principles calculations. The scattering cross section, dominated by the incoherent inelastic contribution from the hydrogen atoms, can be rationalised in terms of the average contributions of different functional groups, thus neglecting their correlation. These results can be used for modelling the total neutron cross sections of complex organic systems like proteins, muscles, or human tissues from a limited number of starting input functions. This simplification is of crucial importance for fine-tuning of transport simulations used in medical applications, including boron neutron capture therapy as well as secondary neutrons-emission induced during proton therapy. Moreover, the parametrized neutron cross sections allow a better treatment of neutron scattering experiments, providing detailed sample self-attenuation corrections for a variety of biological and soft-matter systems.
\end{abstract}

\maketitle

\section{\label{sec:level1} Introduction }

The study of the interaction of neutrons with matter has still to become centennial, yet it has impacted the modern society in a variety of ways, from fission reactors to the creation of isotopes for medical care; from the treatment of cancer to the non-invasive characterization of cultural-heritage artworks; as well as the scientific investigation of the structure and dynamics of condensed-matter systems~\cite{1991_Andreani_RPP,2013_FernandezAlonso_Book}.
In the latter case, special attention should be paid to hydrogen ($^1$H) for it manifests the largest bound scattering cross section (82.03 barn) amongst the elements of the periodic table~\cite{1992_Sears_NN}. For this reason, neutron scattering off hydrogen is particularly relevant, for example, for molecular spectroscopy~\cite{2005_Mitchell_book} and catalysis~\cite{2016_Silverwood_PCCP}, hydrogen storage~\cite{2009_Ramirez-Cuesta_MaterialsToday}, and cultural heritage~\cite{2020_Festa_Sensors}. 

Fermi was the first to model the interaction potential between slow neutrons (energies lower than few eV) and hydrogenous solids~\cite{1936_Fermi_NC}. He explicitly discussed how the total cross sections, $\sigma(E)$, for neutrons with incident energy $E$, would change by approximately a factor of 4 when moving from cold neutrons to epithermal ones.
On the one hand, epithermal neutrons (eV -- keV) have energies much larger than the typical binding energy of hydrogen in a crystal or molecular system, therefore resulting in a Compton-like scattering from an approximately free nucleus~\cite{1986_Gunn_JPC,2005_Imberti_NIMA,
1998_Ricci_JCP,2005_Andreani_AdvPhys,2017_Andreani_AdvPhys,2017_Andreani_Book}. In this case, the scattering is defined as elastic in the neutron + nucleus system, thus requiring the conservation of kinetic energy and momentum of both particles, and the total scattering cross section corresponds to the free-nucleus value, $\sigma_f$.
On the other hand, cold neutrons ($\mu$eV -- meV) impinging on a cold solid sample would bounce off an almost unmovable target. Therefore, as such neutrons do not change their kinetic energy following the interaction, the scattering is elastic in the system composed solely by the neutron, and the total scattering cross section corresponds to the bound-nucleus value, $\sigma_b = \sigma_f(1+m/M)^2$, where $m$ is the mass of the neutron, and $M$ the mass of the nucleus. This corresponds, in the case of hydrogen, to $\sigma_b\simeq 4\sigma_f$.

In the intermediate (thermal) energy region, the scattering is elastic in neither of the two cases aforementioned, for part of the neutron kinetic energy can be transferred to rotational or internal vibrations of a molecule, or lattice and to phonon modes in a crystal, in a regime generally referred to as inelastic neutron scattering~\cite{2005_Mitchell_book}. For crystalline samples, the cross section also depends on the structure via the presence of Bragg edges. Moreover, while the picture drawn by Fermi holds exactly for solids at low temperatures, the total scattering cross section at energies lower than tens of meV can be higher than $\sigma_b$ in a solid at room or higher temperature, following the population of the Stokes and anti-Stokes transitions related to the vibrational modes, according to the Maxwell-Boltzmann statistics. In a liquid, furthermore, contributions from diffusion motions can increase additionally $\sigma(E)$ at values of $E$ below some meV. 

Tabulated Thermal Cross Sections (TCS) are available for just a handful of systems~\cite{1967_Koppel_GASKET,1994_MacFarlane_LANL,2017_MacFarlane_LANL}. 
In order to match the level of detail in modern Monte Carlo nuclear transport codes, the possibility to calculate TCS from simplified models~\cite{1985_Granada_PRB}, from molecular dynamics~\cite{2013_damian_JCP}, as well as from {\it ab initio} calculations~\cite{2004_Hawari_PHYSOR,2014_Hawari_NDS,2017_Cai_NewJPhys,2017_Wormald_EPJConf,2019_Al-Qasir_AnnNuclEn,2019_Cheng_JCTC,2020_Cheng_JCTC} has been a topic of recent development.
Despite the important effort redirected into the theoretical calculation of TCS, the experimental data available for comparison and validation of the models is quite scarce. Moreover, new experimental investigations, aimed to update decades-old data, are clearly needed, as recently demonstrated for the case of para-hydrogen~\cite{2015_Grammer_PRL}. In this context, a sustained experimental programme has been started at the VESUVIO spectrometer~\cite{2017_Romanelli_MST,2020_Robledo_NIMA,2019_Ulpiani_RSI} at the ISIS Neutron and Muon Source (UK)~\cite{ISIS_website}, to investigate TCS of alcohols~\cite{2017_RodriguezPalomino_NIMA}, organic systems~\cite{2019_Capelli_JAC}, water~\cite{2018_Andreani_RNC,2020_Marquez_Damian_EPJ} and other materials used to moderate neutrons~\cite{2018_Romanelli_NIMA,2020_Cantargi_EPJ,2020_Skoro_EPJ,2020_Palomino_NIMA}. The unprecedentedly broad energy range for incident neutrons available at the instrument~\cite{2017_Romanelli_MST}, spanning from a fraction of meV to tens of keV, allows a complete characterization of total neutron cross sections, from cold to epithermal neutrons. Moreover, the broad energy range provides an accurate and self-consistent way to normalize the experimental spectra for those samples whose density is more difficult to determine~\cite{2020_Robledo_NIMA}, such as powders or samples experiencing {\it in situ} adsorption~\cite{2019_Romanelli_JPCC} or phase transitions. 

In this framework, the measurement and calculation of TCS of a vast set of materials are challenging tasks, owing to the several non-trivial dependencies on the molecular structure, dynamics, and thermodynamic temperature. When the TCS calculation is applied to human tissues or muscles for applications such as Boron Neutron Capture Therapy (BNCT)~\cite{2017_ramos_EPJ,2015_ramos_ARI} as well as to study secondary neutrons-emission induced during proton therapy~\cite{2019_Ytre_SR,2017_Lee_Pone}, the possibility to reconstruct the cross section of large proteins becomes challenging. In fact, the 20 basic amino acids can combine to form tens of thousands~\cite{2002_Adkins_MolCelProt} up to several billions of proteins~\cite{2013_Smith_NatMethods}, making the task of either calculate or measure the entire set of related TCS unrealistic.    

Here we provide an experimental determination of the total TCS of the twenty amino acids, as a function of the neutron energy from a fraction of meV to tens of keV, together with first-principles calculations based on the incoherent approximation, thus particularly suitable for hydrogen-containing materials. Given such unprecedentedly consistent set of experimental data, we attempt a rationalization of the TCS of amino acids and, by extension, of proteins, as a set of average contributions of independent functional groups, such as CH$_n$, NH$_n$, and OH. The process of average over the amino acids allows to consider the functional groups as independent, and to obtain the final results directly by adding the different contributions from different functional groups, neglecting their correlations. We will refer to this procedure as the Average Functional Group Approximation, AFGA. This task would allow the replication of complex TCS, {\it e.g.,} for biophysical and medical applications, having just a handful of initial parameters. 

\section{Materials and Methods}
\subsection{\label{sec:level2} Neutron experiment}

Neutron-transmission experiments were performed at the VESUVIO spectrometer at the ISIS Neutron and Muon Source. All the 20 amino acids were commercially available in their L-form from Sigma Aldrich~\cite{Aldrich_website} as anhydrous powders. Samples were loaded as received in circular flat containers with Nb faces perpendicular to the direction of the neutron beam, and Al spacers defining the sample volume. The latter corresponded to a cylindrical shape with thickness of either 1 or 2 mm for all samples, and a circular area facing the beam with diameter 5 cm, thus covering the entire circular beam profile~\cite{2017_Romanelli_MST}.
All samples were measured at the temperature of 300 K within the instrument's closed-circuit refrigerator. For each sample, data were collected for about 0.5--1.5 hours, corresponding to an integrated proton current of 90--270 $\mu$Ah within the ISIS synchrotron. The values of mass, thickness and integrated proton current are reported in Table~\ref{tab:info}, together with the molecular stoichiometry. Time-of-flight spectra of incident neutrons not interacting with the sample were obtained using the standard GS20 $^6$Li-doped scintillator available at the instrument, as well as the newly installed double thick Gas Electron Multiplier (GEM) detector~\cite{2021_Cancelli_JInst}. The latter was positioned between the sample position and the GS20 monitor, at a distance from the moderator of 12.60 m, as calibrated at the beginning of the experiment using the VESUVIO incident foil changer~\cite{2020_Robledo_NIMA}. The GS20 transmission monitor and the sample positions correspond to 13.45 m and 11.00 m, respectively~\cite{2017_Romanelli_MST}. It is important to stress that, for most of the samples investigated in the present experiment, transmission spectra from both the GS20 monitor and the GEM detector were available at the same time, and provided concurrent measurements of the neutron cross section of the same sample using two independent equipments. Given the distances of both detectors, the counts due to multiple scattering events within the sample are completely negligible as compared to experimental errors bars~\cite{2020_Robledo_NIMA}.

Moreover, the GEM detector allowed more precise measurements over the extended energy range available at the instrument, down to 0.6 meV, corresponding to the so-called empty neutron pulse at the Target Station 1 at ISIS~\cite{2017_Romanelli_MST}, as compared to 3 meV of the GS20 detector. This is a consequence of the lower sensitivity to $\gamma$-rays in the GEM compared to scintillators like the traditional GS20 monitor.

\begin{table*}

	\begin{center}
		\begin{tabular}{l|cccccc|cc|ccc|c} 
			\hline
Amino Acid & Formula      &CH$_3$&CH$_2$&CH& 	Zw.$^+$& Other & $\sigma_{f} $[barn]&$\%$H & M [g] & d [mm] & Q [$\mu$Ah] & Structure\\ 
\hline\hline
Leucine   & C$_6$H$_{13}$NO$_2$ 	&2&1&2& 	NH$_3$& -- &      312.08 & 85.3 & 1.63 & 1.00 & 270 &Ref.~\cite{2016_binns_ACb} \\ 
\hline
Isoleucine & C$_6$H$_{13}$NO$_2$    &2&1&2&     NH$_3$& -- &      312.08  & 85.3 & 3.89 & 2.00 & 180 &Ref.~\cite{1996_gorbitz_ACc} \\
\hline
Valine & C$_5$H$_{11}$NO$_2$ 	    &2&--&2& 	NH$_3$& -- &      266.40  & 84.6 & 3.41 & 1.00 & 270 &Ref.~\cite{1996_dalhus_ACS} \\
\hline
Methionine & C$_5$H$_{11}$NO$_2$S   &1&2&1&		NH$_3$&  S &      267.36  & 84.3 & 2.97 & 2.00 & 90 &Ref.~\cite{2016_gorbitz_IUCrJ}\\
\hline
Lysine & C$_6$H$_{14}$N$_2$O$_2$    &--&4&1&     NH$_3$&  NH$_2$ &      342.58  & 83.7 & 3.60 & 1.00 & 270 &Ref.~\cite{2015_williams_AngChemie} \\
\hline
Threonine & C$_4$H$_9$NO$_3$ 	    &1&--&2&    NH$_3$&  OH &      224.46  & 82.1 & 5.02& 2.00 & 90 &Ref.~\cite{1997_janczak_ACc} \\ 
\hline
Alanine & C$_3$H$_7$NO$_2$ 	        &1&--&1& 	NH$_3$& -- &      175.03 & 81.9 & 4.35 & 2.00 & 1080 &Ref.~\cite{wilson2005_wilson_NJC} \\ 
\hline
Cysteine & C$_3$H$_7$NO$_2$S 	    &--&1&1& 	NH$_3$&  SH &      176.00 & 81.5 & 5.30 & 2.00 & 180 &Ref.~\cite{1997_gorbitz_ACc} \\
\hline
Serine & C$_3$H$_7$NO$_3$ 		    &--&1&1&	NH$_3$&  OH &      178.78 & 80.2 & 6.02 & 2.00 & 90 &Ref.~\cite{2005_bolyreva}  \\ 
\hline
Proline & C$_5$H$_9$NO$_2$ 		    &--&3&1& 	NH$_2$& -- &     225.44 & 80.1 & 2.25 & 1.00 & 180 &Ref.~\cite{2018_koenig_ACe} \\ 
\hline
Glycine & C$_2$H$_5$NO$_2$ 		    &--&1&--&   NH$_3$& -- &    129.35 & 79.2 & 4.54 & 2.00 & 270 &Ref.~\cite{2002_drebushchak_ACe} \\
\hline
Arginine & C$_6$H$_{14}$N$_4$O$_2$  &--&3&1&   NH$_2$& (NH$_2$)$_2$,  NH &   362.63  & 79.1 & 3.46 & 1.00 & 270 &Ref.~\cite{2012_courvoisier_CC} \\
\hline
Glutamic Acid & C$_5$H$_{9}$NO$_4$ &--&2&1&	NH$_3$&  OH&   232.93  & 79.1 & 4.16 & 2.00 & 270  &Ref.~\cite{2016_ruggiero_JPCa}\\
\hline
Phenylalanine & C$_9$H$_{11}$NO$_2$ &--&1&6&    NH$_3$&&  285.30  & 79.0 & 6.28 & 2.00 & 90 &Ref.~\cite{2019_cuppen_CGD} \\
\hline
Glutamine & C$_5$H$_{10}$N$_2$O$_3$ &--&2&1&    NH$_3$& NH$_2$&  259.68 & 78.9 & 2.89 & 2.00 & 450 &Ref.~\cite{2001_wagner_JMS} \\
\hline
Tyrosine & C$_9$H$_{11}$NO$_3$ 	    &--&1&5&    NH$_3$& OH&  289.00 & 77.9 & 2.12 & 2.00 & 180 &Ref.~\cite{1973_frey_JCP} \\
\hline
Aspartic Acid & C$_4$H$_{7}$NO$_4$  &--&1&1&    NH$_3$& OH &  187.23  & 76.6 & 5.05 & 2.00 & 270 &Ref.~\cite{2007_bendeif_ACc} \\ 
\hline
Asparagine & C$_4$H$_{8}$N$_2$O$_3$ &--&1&1&    NH$_3$& NH$_2$&   214.00  & 76.6 & 5.39 & 2.00 & 180 &Ref.~\cite{2007_yamada_ACe} \\
\hline
Tryptophan & C$_{11}$H$_{12}$N$_2$O$_2$ &--&1&6&  NH$_3$& NH &      325.23  & 75.6 & 3.05 & 2.00 & 270 &Ref.~\cite{2012_gorbitz_ACb} \\
\hline
Histidine & C$_6$H$_{9}$N$_3$O$_2$   &--&1&3& 	 NH$_3$& NH &      250.20  & 72.7 & 4.69 & 2.00 & 270 &Ref.~\cite{1972_madden_ACb} \\
\hline
\end{tabular}
	\end{center}
	\caption{Chemical formula and decomposition of L-amino acids, together with the total value of the free scattering cross section per formula unit, $\sigma_{f}$, as well as the relative amount of hydrogen atoms per formula unit. The amino acids are generally found in their Zwitterion form, the cation is reported in the column labelled Zw.$^+$, and the formula unit is completed by adding the anion, $COO^-$. For each sample are reported the values of mass ($\pm$ 0.01 g) and thickness ($\pm$ 0.05 mm) together with the integrated proton current. The reference providing the crystal structure used in the {\it ab initio} simulations is also reported in the last column.}
	\label{tab:info}
\end{table*}

Transmission spectra were obtained, as a function of the incident neutron energy, using the Beer-Lambert law, as
\begin{equation}
T(E)=\alpha \frac{S(E)}{C(E)}=\exp\left(-n\sigma(E)d\right),
\label{eq:transmission}
\end{equation}
where $S(E)$ and $C(E)$ are the spectra corresponding to sample inside the container and empty container, respectively. Moreover, $n$ is the number density of molecules inside the sample volume, $\sigma(E)$ is their energy-dependent total cross section, and $d$ the thickness of the container. Finally, $\alpha$ is a normalization factor taking into account the different durations of the measurements with and without the sample. In the case of the GS20 monitor, $\alpha$ is an energy-dependent normalization provided by the measurement of the incident neutron beam by the instrument monitor at 8.57 m from the moderator, before the sample position. In the case of the GEM detector, both $S(E)$ and $C(E)$ were normalized to the number of proton pulses counted by the detector electronics. However, as a slight fluctuation of the efficiency with time is known to affect the GEM detector~\cite{2021_Cancelli_JInst}, of magnitude ca. 2\% with no dependence upon the neutron energy,  the resulting transmission spectra were scaled so as to overlap to the GS20 spectra for epithermal neutrons.

\subsection{Calculations of thermal cross sections}
Within the incoherent approximation~\cite{1958_Sjolander_AF,1970_parks_book,2017_Cai_NJP}, the double differential scattering cross section can be expressed as a sum of single-particle contributions weighted by the sum of coherent and incoherent bound scattering cross sections~\cite{1992_Sears_NN} for each isotope $j$, $\sigma_{b,j}$, namely 
\begin{equation}
\frac{d^2\sigma}{d\mu dE'}= \sqrt{\frac{E'}{E}} \sum_jN_j\frac{\sigma_{b,j}}{2}S_j(\vec{Q},\omega),
\label{eq.1}
\end{equation}
where $\mu$ is the cosine of the scattering angle, $E$ and $E'$ are the initial and final neutron energies, $N_j$ represents the stoichiometry of the sample, and  $S_j(\vec{Q},\omega)$ is known as scattering law (or scattering kernel) as function of the energy, $\hbar\omega=E-E'$ and momentum, $\vec Q$, transfers.
This approximation is particularly suitable for hydrogen-rich materials, as in the present case, for the incoherent scattering contribution from hydrogen sums up to ca. 99\% of the total. Moreover, we adopt the powder-average approximation for non-oriented samples, thus we will consider the scattering law as a function of the modulus of the momentum transfer, $Q$, but not of its direction.
 
$S_j({Q},\omega)$ is made explicit by defining the Fourier transform of the intermediate self-scattering function. Sjolander showed that, in a harmonic cubic Bravais lattice and within the Gaussian approximation, the scattering law can be expressed as~\cite{1958_Sjolander_AF,1970_parks_book}
\begin{equation}
S_j(Q,\omega) 
= e^{-2W_j} \sum_{n=0}^\infty \frac{(2W_j)^n}{n!} \,\, H_n(\omega),
\label{exp}
\end{equation}
where
$H_0(\omega) =  \delta(\omega)$, \cite{2017_Cai_NewJPhys}
\begin{equation}
H_1(\omega) = \frac{g_j(\omega)}{\hbar\omega\, \gamma_j(0)}\frac{1}{2}\left[  \coth \left( \frac{\hbar\omega}{2k_BT} \right) +1 \right]
\end{equation}
and
\begin{equation}
H_{n>1}(\omega) = \int H_1 (\omega')H_{n-1}(\omega - \omega^\prime) \,  d\omega^\prime  \, .
\label{H_n}
\end{equation}
This procedure describes the so-called Multi-Phonon Expansion (MPE), where the $n$-th term represents the contribution from the scattering process involving $n$ phonons. Thus, the first term gives the elastic cross-section; the next term gives the cross-section for all one-phonon processes in which, in turn, each phonon is excited; and so on. The function $g_j(\omega)$ represents the unit-area normalized Vibrational Density of States (VDoS) of a given nucleus, such that $g_j(\omega) d\omega$ is the fraction of the normal modes whose frequencies lie in the range between $\omega$ and $\omega + d\omega$. 
Moreover, in Eq.~\ref{exp}, 
\begin{equation}
\gamma_j(0)=\int_{0}^{\infty} \frac{g_j (\omega)}{\hbar\omega} \coth\left(\frac{\hbar\omega}{2k_BT}\right)d\omega ,
\label{gamma0}
\end{equation}
is also used to define the so-called Debye-Waller factor, $2W_j$, \cite{1979_squires_book} with the relation
\begin{equation}
2W_j= \langle(\vec{Q}\cdot\vec{u_j})\rangle= \frac13 Q^2 \langle u^2_j\rangle = E_{R,j}\gamma_j(0),
\end{equation}
where $E_{R,j}=\hbar^2Q^2/2M_j$, $\langle u^2_j\rangle$ is the mean square displacement, and the second equality holding within the isotropic approximation. 

In principle, the MPE can be calculated up to any value of $n$, yet its numerical evaluation becomes progressively more time-consuming. In fact, as the incident energy increases, an increasingly large number of phonons is likely to be involved in the scattering process. However, in the epithermal region the neutron scattering cross section can be interpreted in the Impulse Approximation (IA)~\cite{1989_Mayers_PRB_ISE} and the scattering law can be expressed as
\begin{equation}
S_{IA,j}(Q,\omega)= \frac{\exp\left(-\frac{(\hbar\omega-E_{R,j})^2}{4E_{R,j} k_BT_j}\right)}{\sqrt{4\pi E_{R,j} k_BT_j}},
\label{IA}
\end{equation}
where $E_{R,j}$ is interpreted, within the IA, as the nucleus recoil energy, and $T_j$ is an effective temperature proportional to the kinetic energy of the scattering nucleus. 
In this work, we have adopted the transition energy between the MPE and the IA equal to 3 eV. In particular, above this energy and only for the evaluation of the total scattering cross section, the value of $T_j$ can be approximated with $T$, the same thermodynamic temperature used in the MPE. Otherwise, at lower values of $E$, substantially higher values of $T_j$ should be used~\cite{2017_RodriguezPalomino_NIMA} in the case of hydrogen well above room temperature (see {\it e.g.,} Ref.~\cite{2014_Moreh_CP,2018_Romanelli_Mantid,2019_Andreani_SCPMA,2020_Andreani_JPCL}) and for heavier masses up to room temperature (see {\it e.g.,} Refs.~\cite{2013_Romanelli_JPCL,2017_Syrykh_JETP,2020_Perrichon_ChemMat,2020_Finkelstein_ChemPhys})

The total scattering cross-section $\sigma(E)$ is obtained by combining Eqs.~\ref{exp} and \ref{IA} with Eq.~\ref{eq.1}, integrating over the values of $\mu$, in the range $[-1,1]$, and over $E^\prime$. The VDoS can be obtained from first-principles computer simulations, as discussed in the next section.
From a numerical point of view, it is important to note that the limit of energy integration is only formally extended to infinity because the integrand differs from zero only in a finite interval up to the cut-off energy $\hbar\omega_{C}=n\hbar\omega_{m}$, with $\hbar\omega_m$ the highest vibrational frequency in the VDoS. Furthermore, each term $H_n(\omega)$ is unit-area normalized. Finally, the total cross section is obtained by adding the neutron absorption cross section. By neglecting any nuclear resonance, not expected for H, C, O, S and N in the range investigated in this experiment, the absorption contribution has a simple form $\sigma_{a,j}(E)=\sigma_{a,j}\sqrt{E_0/E}$, where the values of $\sigma_{a,j}$, corresponding to the absorption at $E_0=25.3$ meV, can be adopted from Ref.~\cite{1992_Sears_NN}.
Therefore,
\begin{equation}
\sigma(E)= \int d\mu \int dE^\prime \, \frac{d^2\sigma}{d\mu dE^\prime} + \sum_j N_j\sigma_{a,j}(E).
\end{equation}

\subsection{Phonon calculations}
\label{sec:mat_meth_DFT}
The VDoS used in the MPE to estimate the TCS was obtained from {\it ab initio} simulations using the Quantum Espresso (QE)~\cite{QE} code, with a 6$\times$6$\times$6 k-grid in the first Brillouin zone of the crystal structure. The pseudopotentials used were taken from Ref.~\cite{pseudo_QE}, while the crystal structures for each amino acids were taken from the Refs. reported in Table~\ref{tab:info}.

A Density Functional Perturbation Theory (DFPT) calculation was run on the optimized geometry, the output corresponded to a collection of eigenvalues of the molecular normal modes, $\omega_v$, and eigenvectors representing each atomic motion for a given normal mode, $\vec e_{v,j}$. The index $v$ spans between 1 and $3N-3$ normal modes, where $N$ is the number of atoms in the unit cell and where we neglected the three lowest-energy translational modes.
The atom-projected VDoS for atom $j$ is obtained as
\begin{equation}
g_{j}(\omega)=\sum_{v} \delta(\omega - \omega _v)  e^2_{v,j} ,
\end{equation}
where $e^2_{v,j}$ is the square modulus of the eigenvector of $j$-nucleus corresponding to the $v$-vibration. In particular, the quantities $\vec e_{v,j} = \sqrt{M_j}\vec u_{v,j}$ are obtained from the dynamical matrix and are associated to the eigenvalue frequency $\omega_v$. Finally, $\vec u_{v,j}$ is the related atomic displacement, expressed as the difference of the time-dependent and mean nuclear positions. 
%By imposing that, for each vibration $v$, the sum of $e^2_{v,j}$ over all atoms is normalized to unity, the total VDoS is normalized to the number of vibrations. 
Additional details regarding the calculation of eigenvectors and eigenfrequencies were reported in Ref.~\cite{2015_togo_SM,2020_Ulpiani_JCP}.

One should note that the {\it ab initio} simulations provide substantial additional information with respect to the experiment, as one can analyse the VDoS for each nucleus in the molecule separately. Such functions can either be averaged over nuclei of the same element, such as H, N, C, S and O in the present case, or they can be averaged for nuclei in the same functional group and over different molecules, such as H in a CH$_3$ methyl group. 

\section{Results}
The experimental neutron transmission spectra of all amino acids were interpreted according to Eq.~\ref{eq:transmission}. In particular, the value of $-\ln(T(E> 3\mbox{ eV}))$ was normalised to the constant free scattering cross section of the molecule,
\begin{equation}
\sigma_{f}=\sum_{j} N_j\,\sigma_{f,j},
\label{normalization}
\end{equation}
where $\sigma_{f,j}$ is the free scattering cross section for isotope $j$, and $N_j$ the related stoichiometry. This is a fundamental tool available at the VESUVIO spectrometer, for it allows a self-consistent normalization of the data not based on the measured sample density,  dependent on the powder grain size and packing fraction, and quite different from the tabulated one for the bulk crystal. The values of the bound cross sections for H, O, C, N, and S were taken from \cite{1992_Sears_NN}, and we considered for each element the values averaged over the naturally abundant isotopes.
\begin{figure}[htbp]%figura1
	\graphicspath{{./figures/}}
	\centering 
	\includegraphics[width=0.50\textwidth]{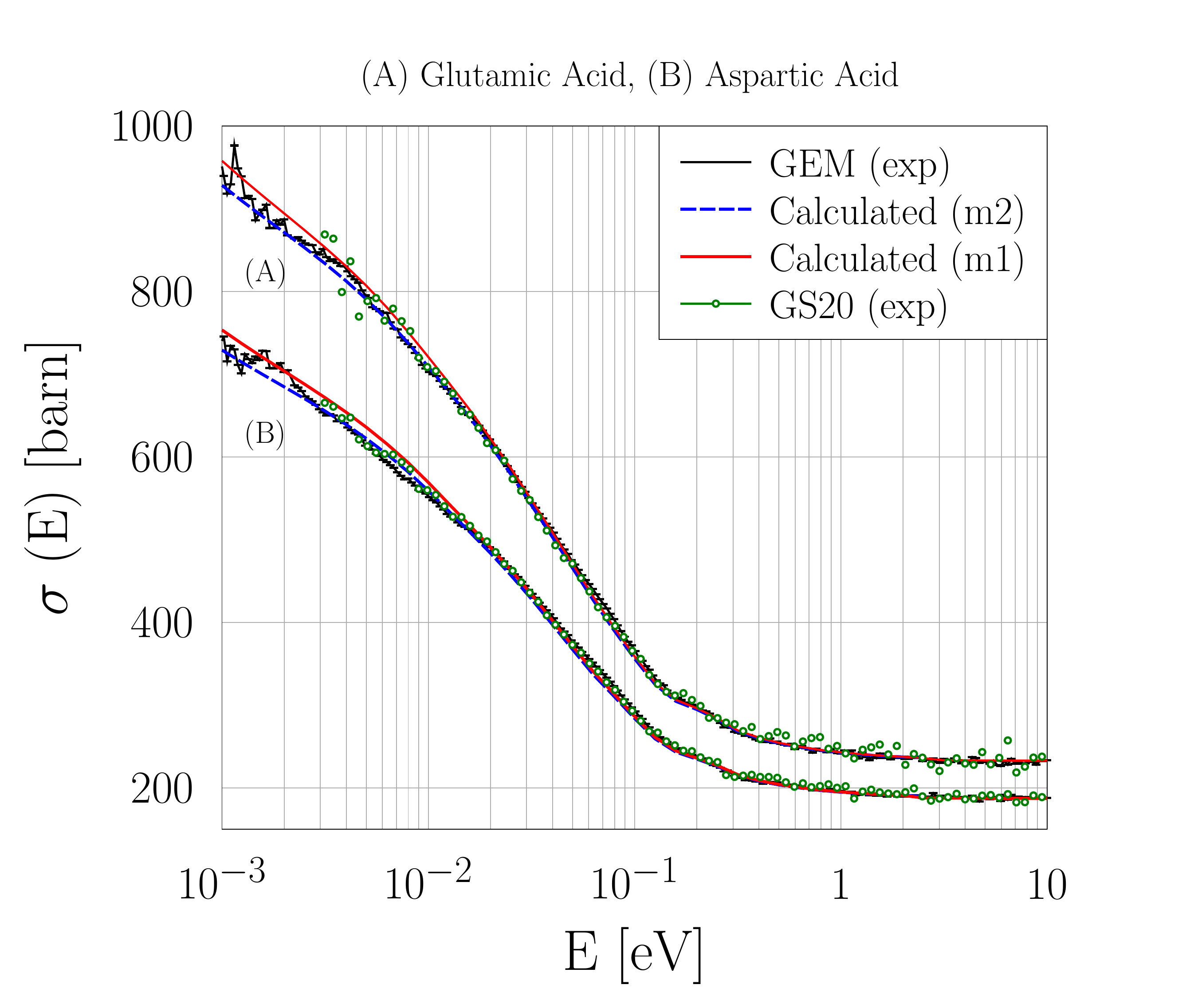}
	\includegraphics[width=0.50\textwidth]{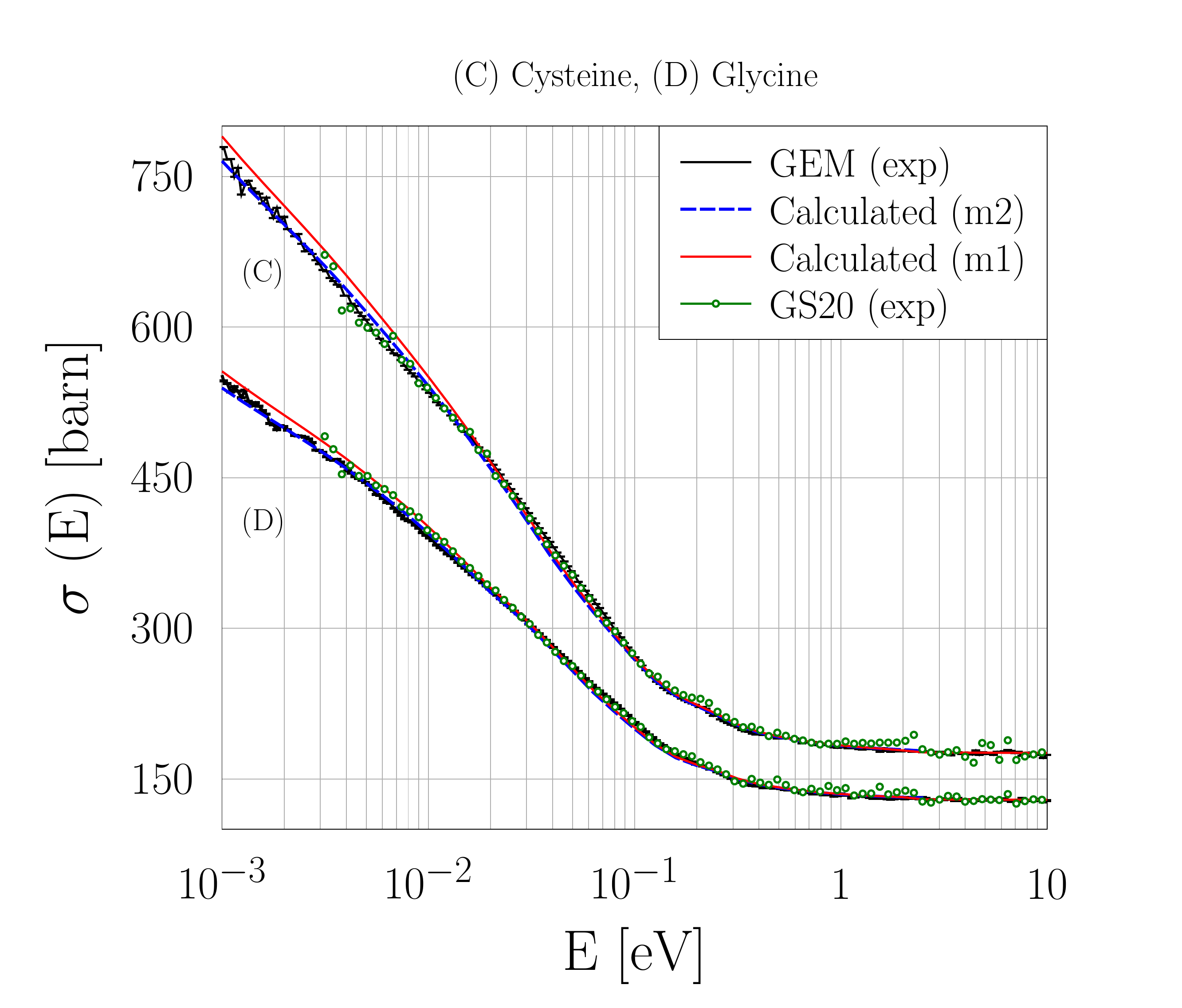}
	\includegraphics[width=0.50\textwidth]{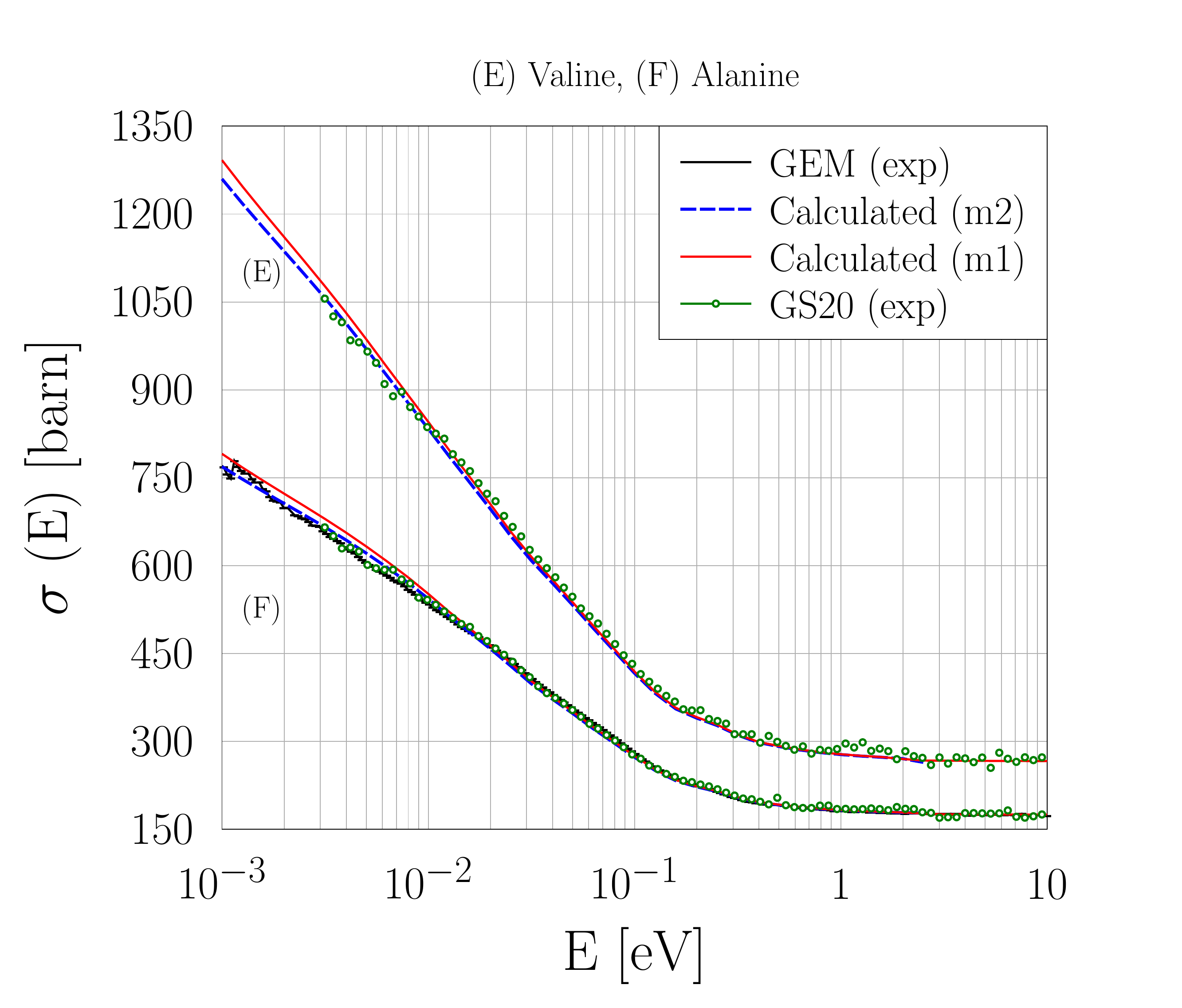}
\caption{\label{fig:best} The total cross section of glutamic acid (A) and aspartic acid (B); cysteine (C) and glycine (D); and valine (E) and alanine (F). For each sample, the experimental data from the GEM (black error bars) and GS20 monitors (green circles) are compared to the calculated spectra using model 1 (red continuous line) and model 2 (blue dashed line). }
\end{figure}

Figure~\ref{fig:best} shows the experimental total cross sections of (from top to bottom) glutamic acid (A) and aspartic acid (B); cysteine (C) and glycine (D); and valine (E) and alanine (F). For each sample, the results obtained using the GS20 and GEM monitors agree, within error bars, over the neutron energy range spanning from 3 meV to 10 eV. For the investigated energies higher than 10 eV, up to 10 keV, all cross sections are found to be constant, as expected. Importantly, the new GEM detector allows a more precise determination of $\sigma(E)$ in two key regions. First, more accurate data are obtained at energies between 0.6 meV and 3 meV, using the empty-pulse at ISIS Target Station 1~\cite{2017_Romanelli_MST} where the spectra collected by the GS20 can be affected by an important environmental gamma background~\cite{Onorati_2018_eVSWorkshop,2020_Onorati_NIMA,2018_ulpiani_eVSWorkshop}, especially for samples with high scattering power. Yet, owing to the lower sensitivity to $\gamma$-rays provided by the GEM detector, the data collected with this new addition are of exquisite quality. The second region where the new detector provides a relevant improvement to the VESUVIO instrument corresponds to epithermal neutrons. Here, the higher efficiency of the detector allows a higher count rate, thus smaller experimental error bars and less scattering of the data points around the value of the free scattering cross section. This is of particular importance for experiments involving powder samples, as in this case, for it improves the normalization procedure based on Eq.~\ref{normalization} and discussed in Ref.~\cite{2020_Robledo_NIMA}.

The experimental data in Figure~\ref{fig:best} were compared to the predictions from the MPE and IA, as discussed earlier. In particular, two models were tested for the scattering contributions: model 1 (m1) applies the MPE to all elements in the molecule; while model 2 (m2) combines the result from the MPE applied to hydrogen only and the values of the free scattering cross sections, $\sigma_{f,i}$ for the elements $i$ other than hydrogen. As one can appreciate from the comparison in Figure~\ref{fig:best}, both models compare extremely well to the experimental data, and they differ only slightly one from the other. In particular, the values from model 1 are found to be, for all amino acids, slightly larger than those from model 2, as expected, and also slightly larger than the experimental data. One should note that the MPE applied to the elements heavier than hydrogen is more sensitive to the lower-energy part of $g_i(\omega)$, where the atom-projected VDoS is more intense. Following the $1/\omega$ dependence in Eq.~\ref{gamma0}, one can expect that a slight error in the calculation of the intensities of $g_i(\omega\to0)$ becomes particularly evident on the final result, and especially at higher values of $T$. It is also important to note that the calculation of $\sigma(E)$ using model 2, {\it i.e.,} applying the MPE only to hydrogen and using $\sigma_{f,i}$ for the other elements, provides a significantly simpler approach and substantially lower computational costs, yet obtaining a very accurate result. Experimental results for the neutron cross section of amino acids were previously presented in Ref.\cite{2017_Voi_INAC} at one value of the incident neutron energy of 50 meV. We note that those results generally underestimate the results from the present investigation, possibly because of the uncertainties on the density and thickness of the powder samples avoided in our normalization procedure~\cite{2020_Robledo_NIMA}.

Finally, we note that all calculated cross sections were obtained by a newly created Python algorithm, based on the formalism explained in Section 2, and prepared in such a way that it can be readily included in the MantidPlot \cite{Mantid_website,2018_Romanelli_Mantid} environment for the reduction and analysis of neutron experiments. In particular, the application of the script developed here would be suitable for the reduction and normalization algorithm presented in Ref.~\cite{2020_Scatigno_JCTC}. While the agreement between our calculations and the experimental results is a powerful benchmark of our algorithm, we performed additional comparisons applying the LEAPR module of NJOY2016~\cite{2017_MacFarlane_LANL} to our VDoS. The results from our algorithm satisfactorily overlap with those obtained using NJOY2016, with a maximum difference in the case of glycine at 1 meV of about 0.7\%, and, therefore are not shown here.

\section{Discussion}
\subsection{Average Functional Group Approximation}
A final, fundamental, comment regarding Figure~\ref{fig:best} is related to the different features and energy dependence of the cross section of each amino acid. As mentioned earlier, this is the result of the different VDoSs that make, in principle, the determination of the cross section of proteins a  challenging task. In particular, pioneering studies using inelastic neutron scattering already showed how the VDoS (thus the double-differential cross section, Eq.~\ref{eq.1}) of isolated amino acids could differ significantly from those of the related dipeptides~\cite{2008_Parker_Spectroscopy}. However, when integrating Eq.~\ref{eq.1} so as to obtain the total cross section, some sharp features and differences vanish. In this framework, we have interrogated our {\it ab initio} simulations so as to extract the contribution to the total cross section from hydrogen atoms in different functional groups within the molecule. 

\begin{figure}[htbp]%figura1
	\graphicspath{{./figures/}}
	\centering 
	\includegraphics[width=0.40\textwidth]{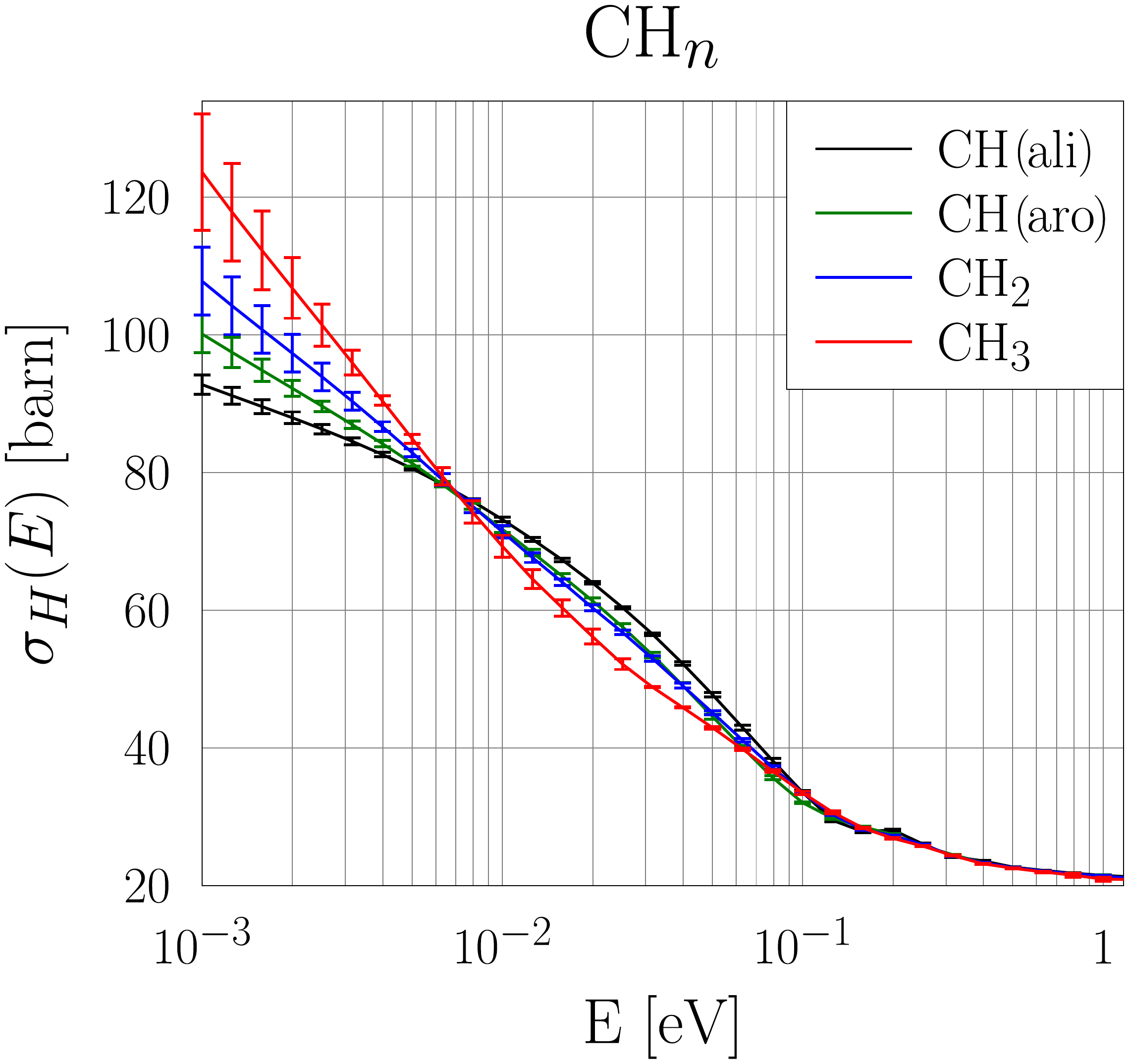}
	
	\includegraphics[width=0.40\textwidth]{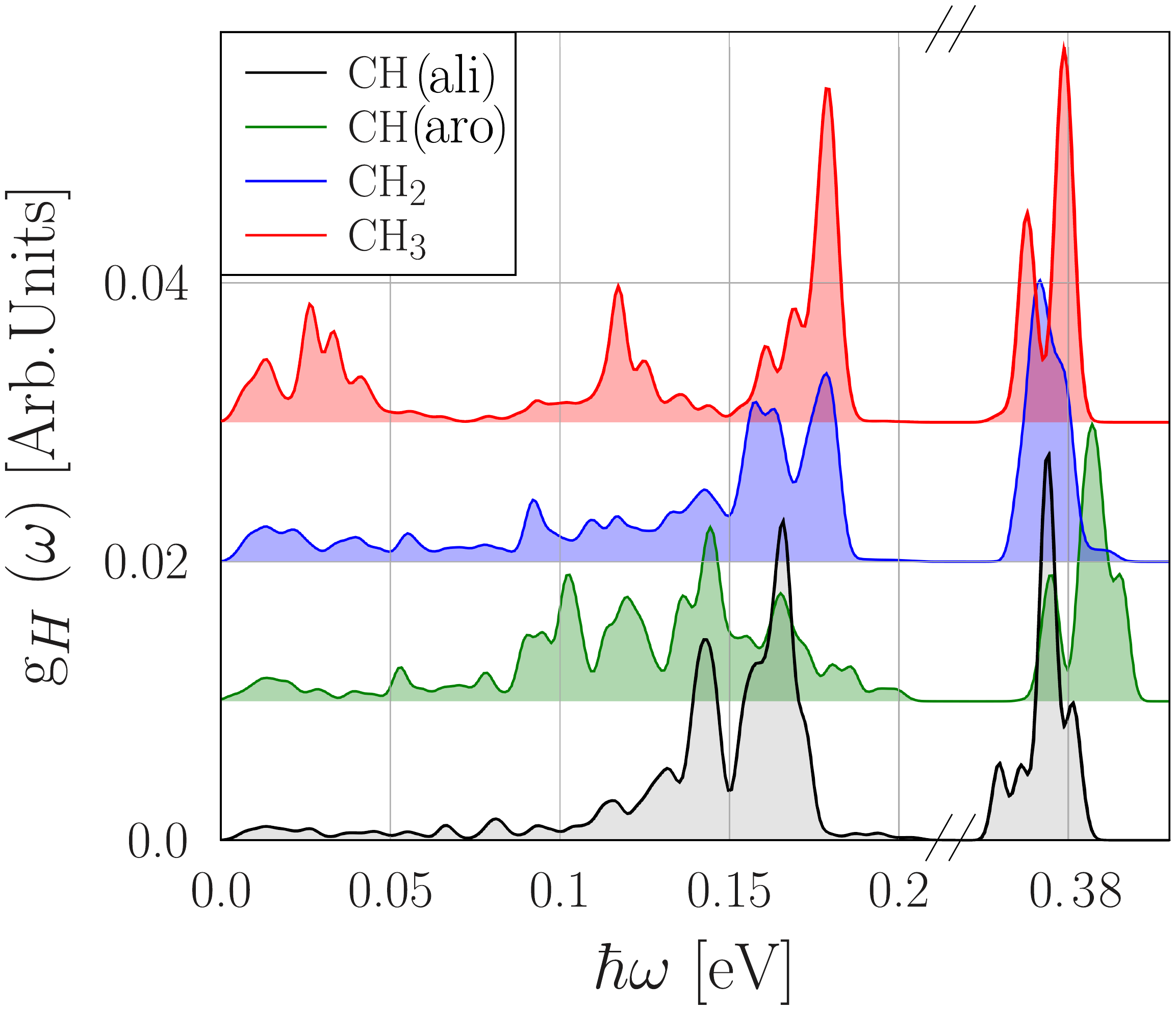}
	\caption{Total cross section (top) and atom-projected VDoS (bottom) of hydrogen in CH$_n$ functional groups, averaged over all amino acids presenting such groups. The VDoSs spectra are shifted in the y-scale for the sake of clarity.}
	\label{fig:funC}
\end{figure}

\begin{figure}[htbp]%figura1
	\graphicspath{{./figures/}}
	\centering 
	\includegraphics[width=0.40\textwidth]{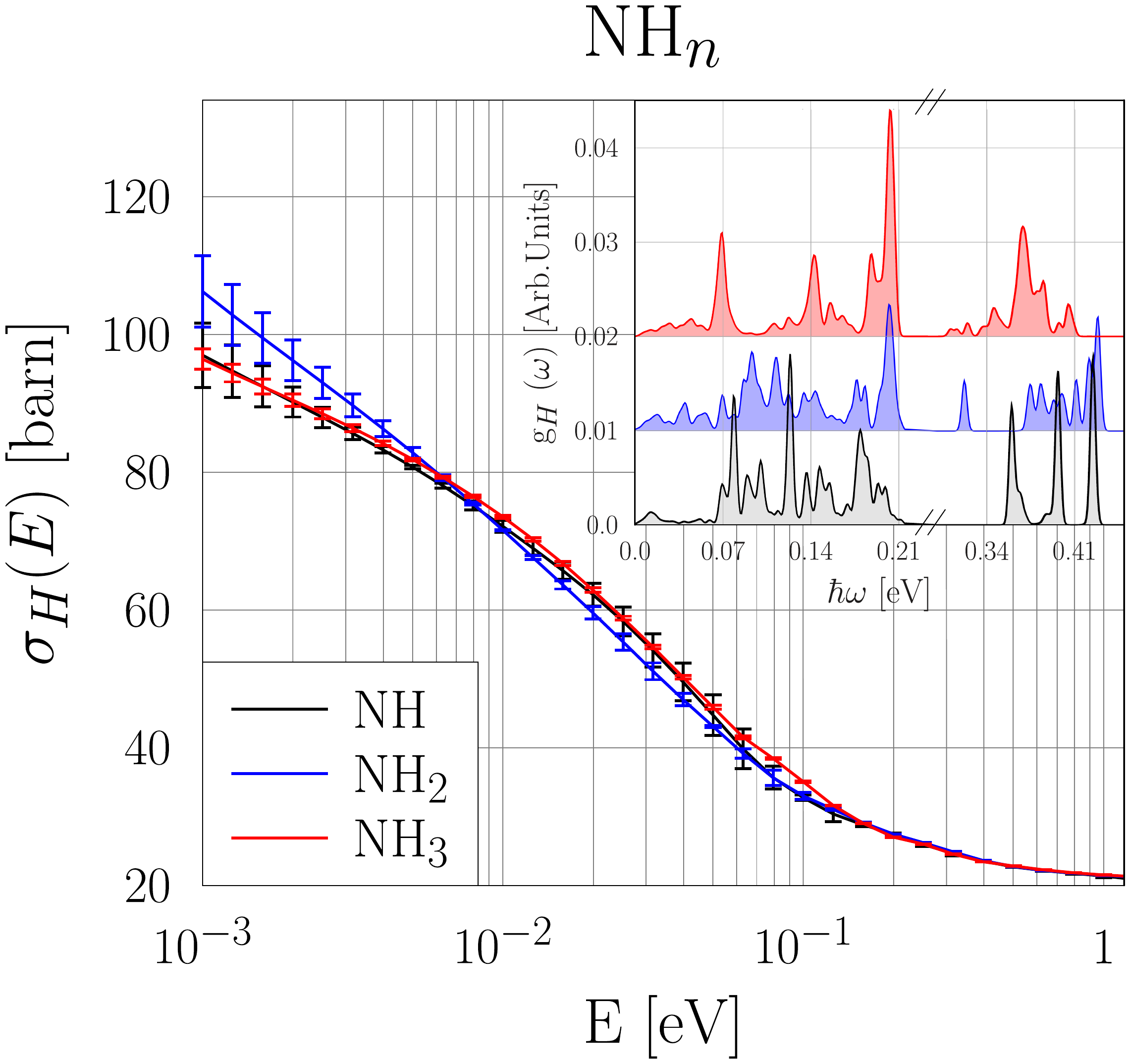}

	\includegraphics[width=0.40\textwidth]{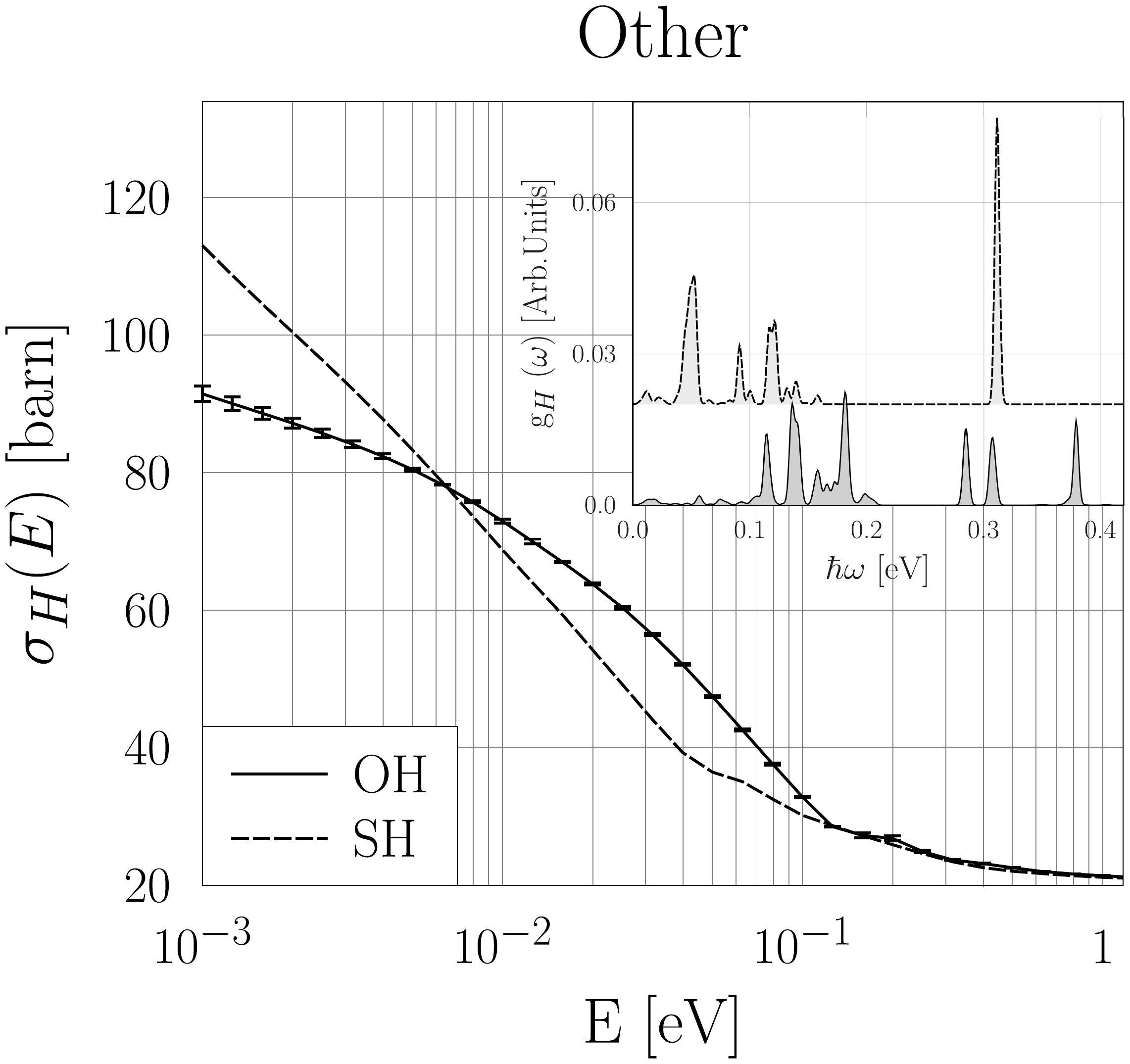}
	\caption{ Total cross section of hydrogen in NH$_n$, OH, and SH functional groups, averaged over all amino acids presenting such groups.
	In the inserts the respectively atom-projected VDoS. }
	\label{fig:funNO}
\end{figure}

Figures~\ref{fig:funC} and~\ref{fig:funNO} show the total cross section of hydrogen in different average functional groups, namely CH$_n$, NH$_n$, OH, and SH. In particular, the molecular structure of each amino acid simulated was interpreted according to the separation in functional groups reported in Table~\ref{tab:info}. The total cross section from hydrogen atoms in the same or different amino acids but belonging to the same type of functional group were averaged, and for each value of the incident neutron energy, a standard deviation associated to this average was defined and corresponds to the error bars in the figures. The fact that signals from the functional groups are found to differ beyond the calculated error bars is an {\it a posteriori} proof that the averaging procedure was meaningful. One can appreciate, for example looking at the top panel of Figure~\ref{fig:funC}, how the average contribution to the total cross section in a methylidyne ($\equiv$CH), methylene ($=$CH$_2$), and methyl groups ($-$CH$_3$) are markedly different one from the other. In particular, within the CH grouping, one can distinguish the contributions from aliphatic CH and aromatic CH functional groups. This has to do with the different VDoSs in the four groups, as shown in Figure~\ref{fig:funC} (bottom). For example, the Stokes and anti-Stokes low-energy excitation and de-excitations corresponding to the CH$_3$ rotor in the methyl group are more easily populated at room temperature, and provide a higher value of the total cross sections at thermal-to-cold neutron energies.  On the other hand, as all VDoSs are normalized to one, the larger cross section of hydrogen in a CH$_3$ group at neutron energies lower than ca. 7 meV has, as a counterpart, a depletion in the region between ca. 7 meV and ca. 90 meV. The opposite is true for the hydrogen in a methylidyne group, while hydrogen in a methylene group has intermediate values between the two previous cases. It is important to note that the average of cross sections over hydrogen atoms participating in same functional group is not identical, in principle, to the cross section obtained from the average of VDoSs of the same hydrogen atoms. However we have checked that the results obtained following these two approaches, in the case of CH, CH$_2$ and CH$_3$, are the same within our numerical accuracy. For this reason, we report in the Supplementary Material the average VDoS of the functional groups considered so as to allow the calculation of the total cross sections at temperatures other then 300 K. 

\begin{figure*}[htbp]%figura1
	\graphicspath{{./figures/}}
	\centering 
	\includegraphics[width=0.48\textwidth]{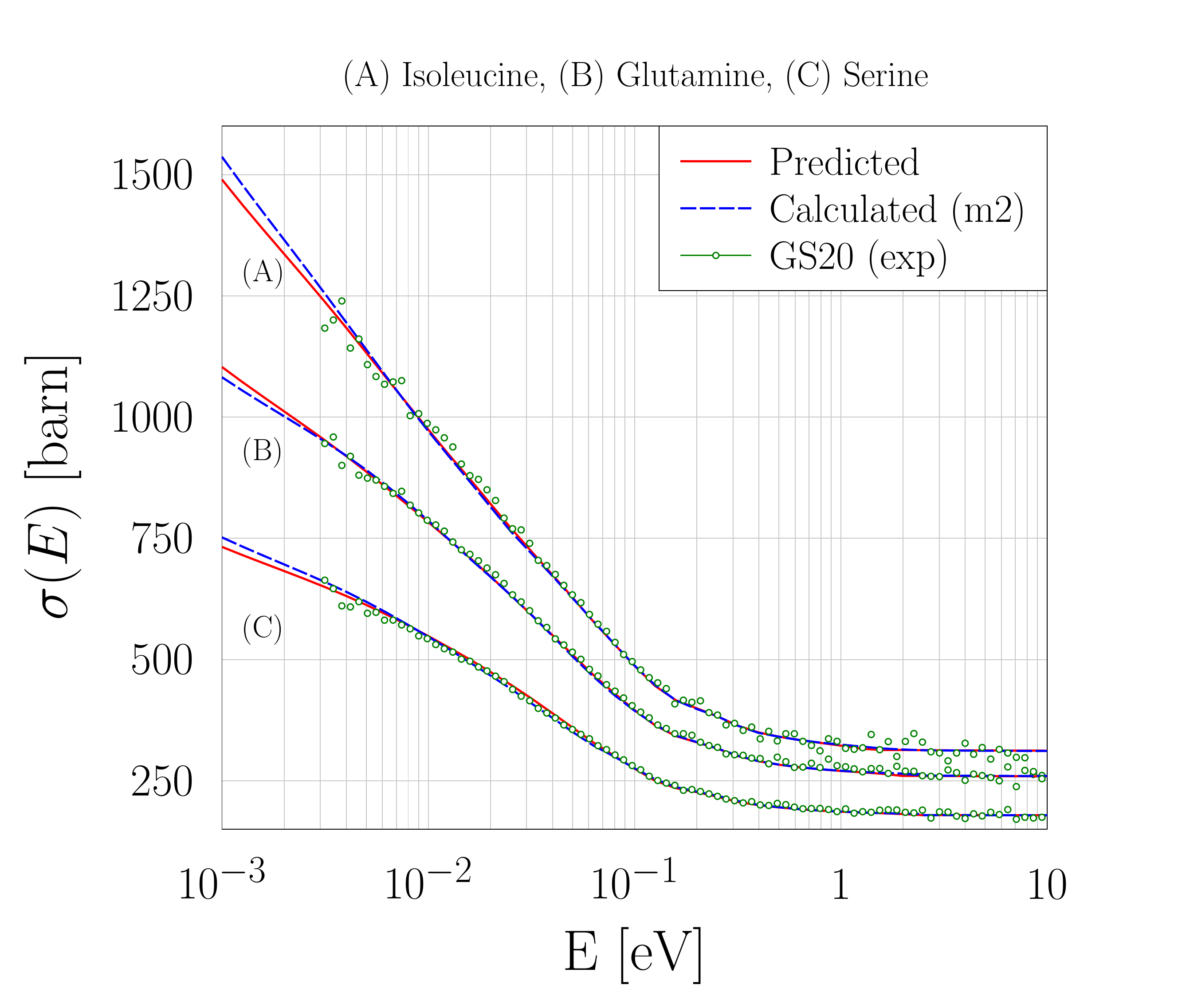}
	\includegraphics[width=0.48\textwidth]{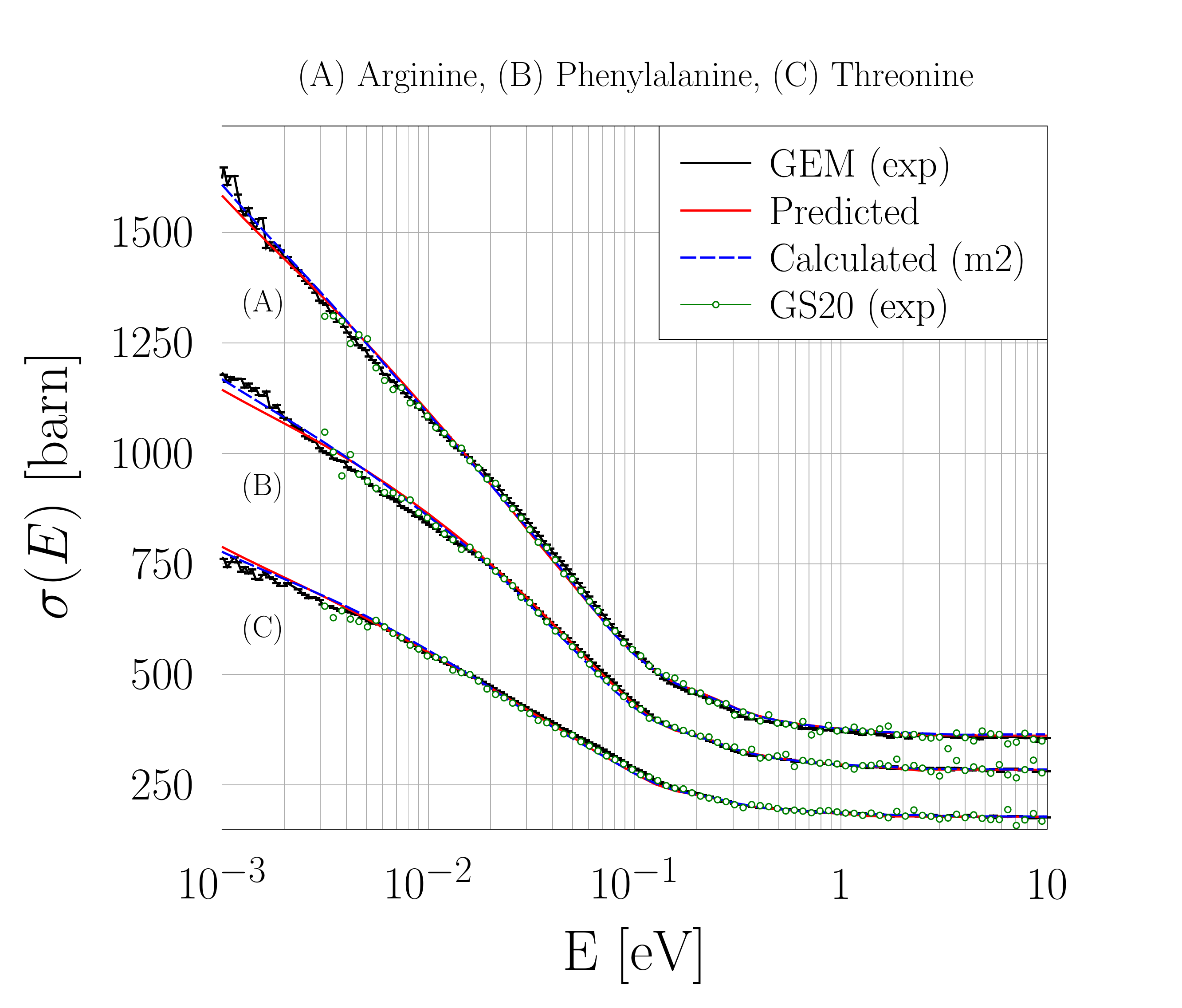}
	\includegraphics[width=0.48\textwidth]{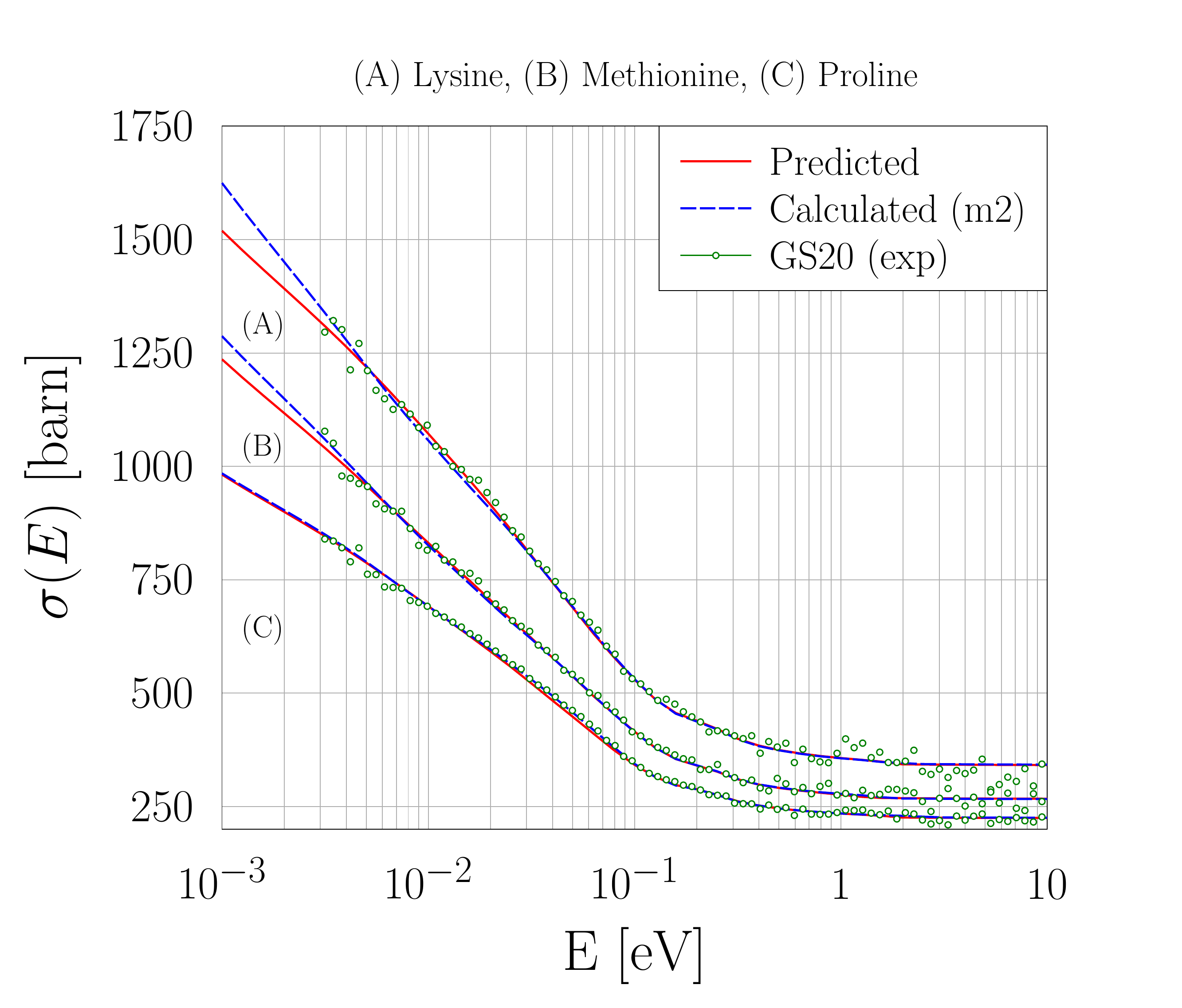}
	\includegraphics[width=0.48\textwidth]{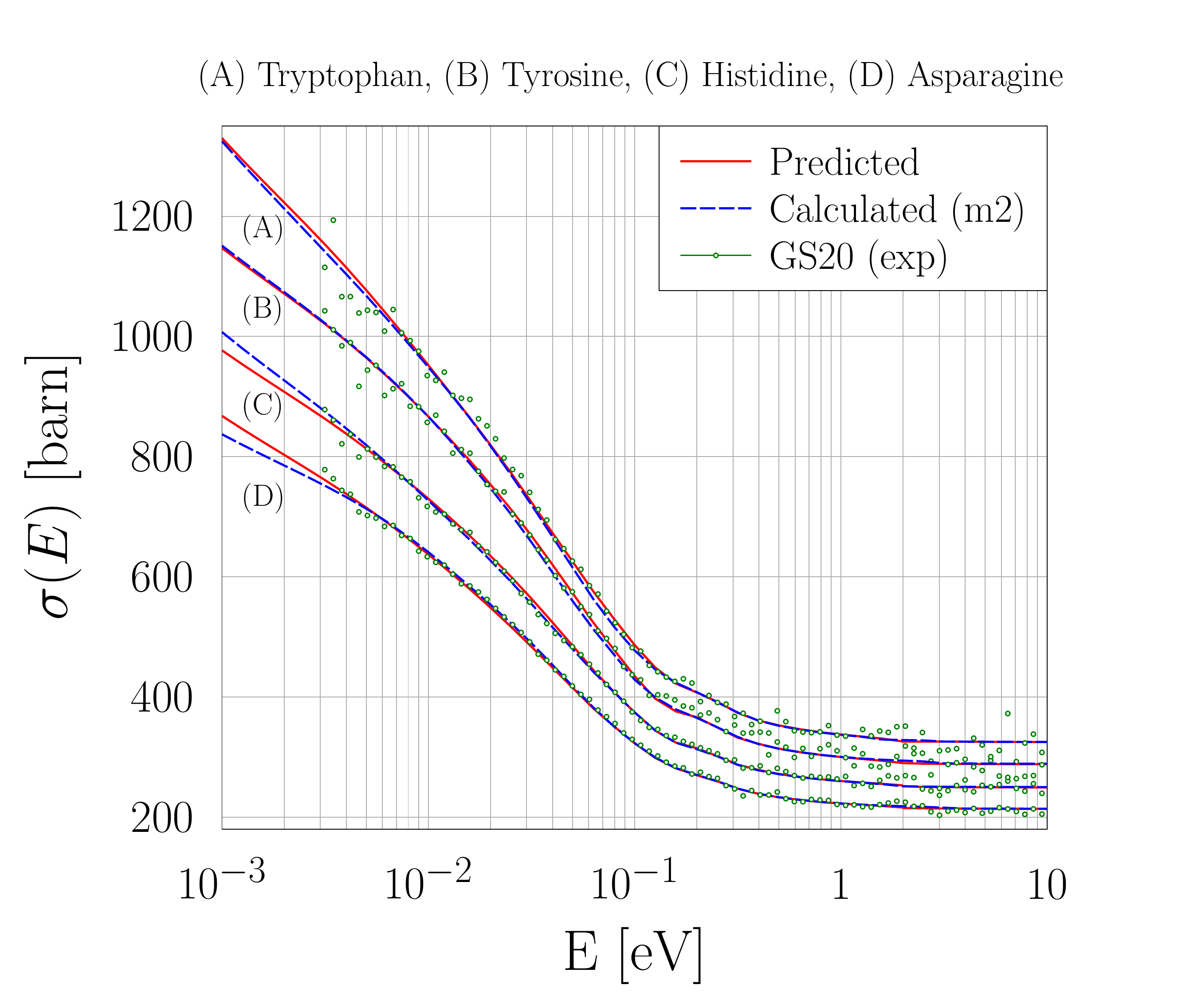}
\caption{\label{fig:results} Total neutron cross section for the 13 amino acids not reported in Figure~\ref{fig:best}. The results obtained using the Average Functional Group Approximation (AFGA, red line) are compared to the experimental results obtained using the GS20 (green circles) and GEM detector (black error bars). The results obtained applying model 2 (m2) to the specific phonon calculation of each molecule are also reported as blue dotted line. }
\end{figure*}

The possibility to express the total cross section of amino acids as average contributions of independent functional groups allows, in principle, the possibility to accurately approximate the cross section of a given protein by the {\it a priori} knowledge of its composition and the set of 9 functions reproduced in Figures~\ref{fig:funC} and~\ref{fig:funNO} and reported in the Supplementary Material from 1 meV to 1 eV . In order to test this approximation, we have compared the experimental cross sections of the 13 remaining amino acids with the predictions based on the AFGA. The results, presented in Figure~\ref{fig:results}, are also compared with the calculations using model 2. In general, the predictions compare extremely favourably to both experimental data and {\it ab initio} calculations based on model 2. Some differences arise for neutron energies lower than few meV, where we find, in the case of lysine, a maximum difference of ca. 5\% at 1 meV between the prediction based on the AFGA and the calculation based on model 2. As we have proved correct the AFGA for a large set of amino acids, one can expect that the same approximation will hold as well for larger proteins. 

\subsection{Application to other organic materials}
\begin{figure}[htbp]
	\graphicspath{{./figures/}}
	\centering 
	\includegraphics[width=0.43\textwidth]{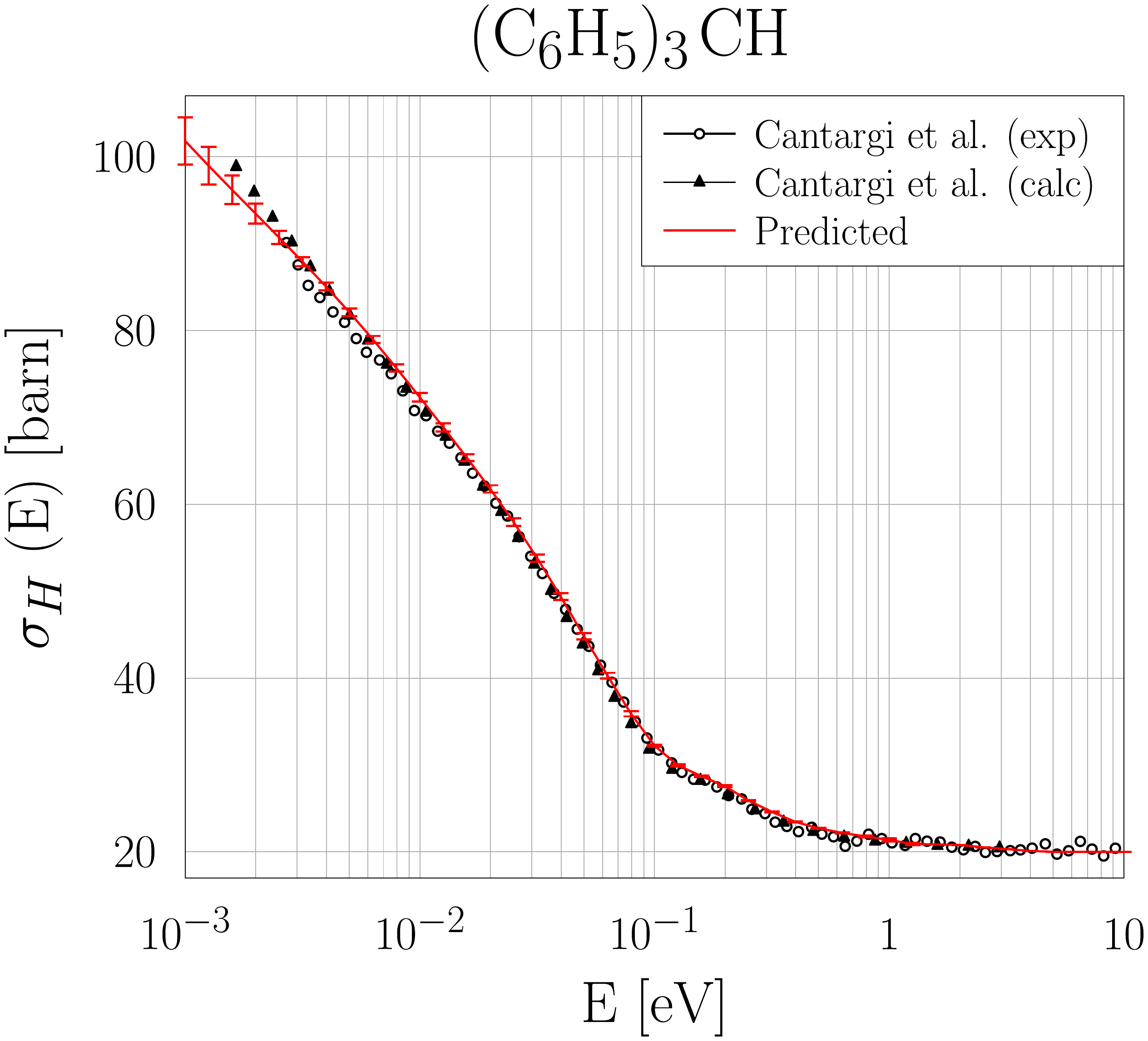}
	
	\includegraphics[width=0.52\textwidth]{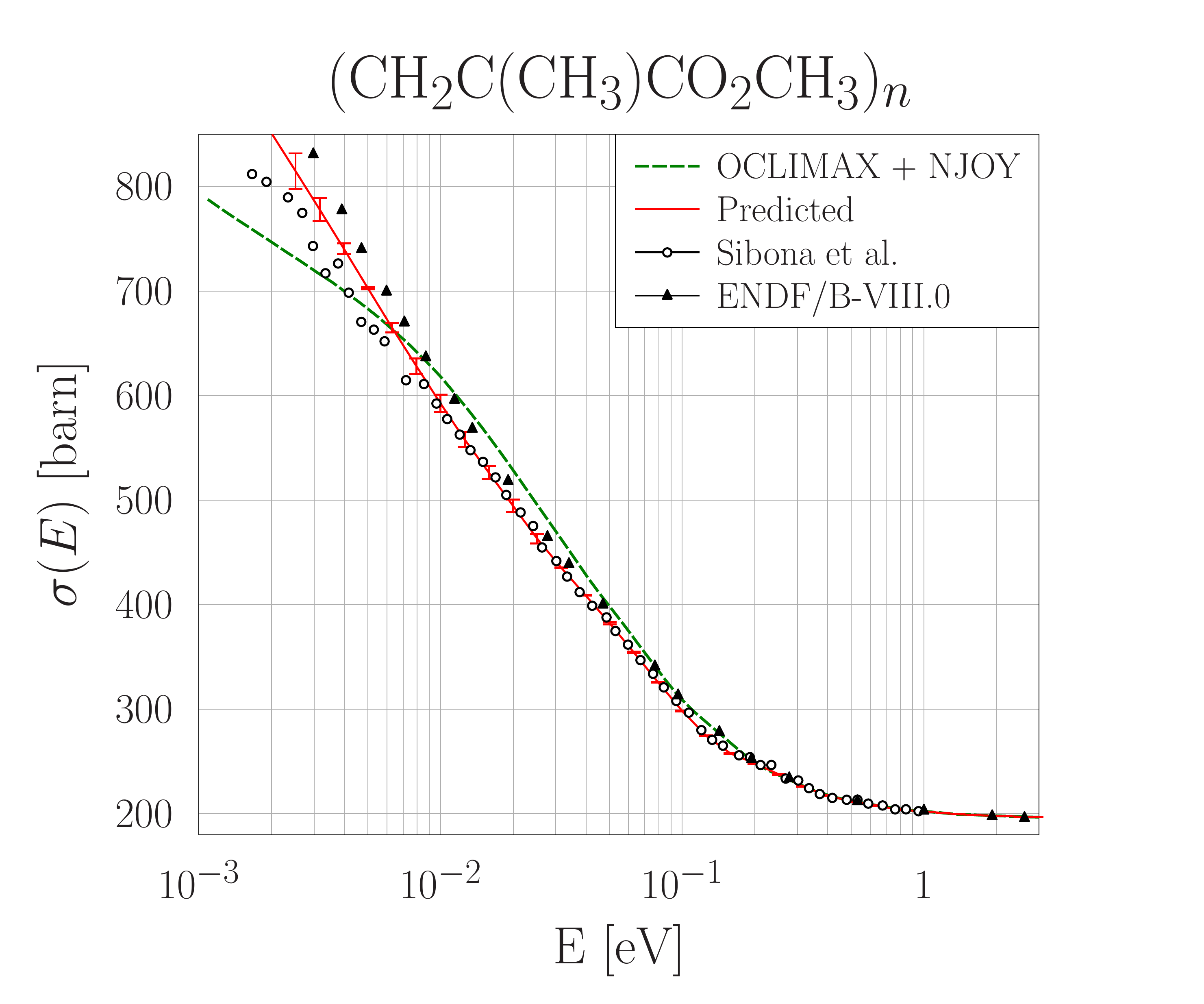}
\caption{\label{fig:comparison}Hydrogen neutron cross section in solid triphenyl-methane (top) from Ref.~\cite{2020_Cantargi_EPJ} and the prediction based on the AFGA. Total neutron scattering cross section of poly-methylmethacrylate or lucite (bottom) from Ref.~\citep{2019_Ramic_ANE} and the prediction based on the AFGA.}
\end{figure}
Triphenylmethane (C$_{6}$H$_{5}$)$_3$CH is a material of great potential interest as neutron moderator. This molecular compound is composed by three phenyl groups connected through a central carbon atom. In the top panel of Figure~\ref{fig:comparison}, it is shown the comparison between the experimental hydrogen cross section of triphenylmethane, measured on VESUVIO from Ref.~\cite{2020_Cantargi_EPJ}, and the predicted spectrum using the average cross section from aromatic CH functional group defined in our AFGA model. In the figure it is also reported the calculation, from Ref.~\cite{2020_Cantargi_EPJ}, from NJOY2016 Nuclear Data Processing system using as input the VDoS obtained by DFT simulations. We note that the agreement of our results compared to experimental data and simulation is excellent, providing a successful test of AFGA also for molecular systems different from amino acids. 

The bottom panel of Figure~\ref{fig:comparison} shows the comparison between the neutron scattering cross section of poly-methylmethacrylate, or lucite (C$_5$O$_2$H$_{8})_n$, from Ref.~\citep{2019_Ramic_ANE} and the prediction based on the AFGA. To construct the total cross section based on the AFGA, we considered lucite as composed by non-interacting chains of CH$_2$C(CH$_3$)CO$_2$CH$_3$ units. 
The related DFT calculations from Ref.~\citep{2019_Ramic_ANE} were performed using CASTEP~\cite{2005_CASTEP} and the output files were processed with OCLIMAX~\cite{2019_Cheng_JCTC} to calculate the scattering law, then with NJOY2016~\cite{2017_MacFarlane_LANL} to calculate the neutron scattering cross section using a procedure tested in other articles, {\it e.g.,} Ref.~\citep{2018_Ramic_ANE}. In Figure~\ref{fig:comparison} (bottom), the experimental data measured by Sibona et al.~\citep{1991_Sibona_ANE} are also reported, as well as the cross section tabulated in the ENDF/BVIII.0~\citep{2018_Brown_NDS} library. A very good agreement between the experimental data and the reconstruction using our AFGA model has been found over the entire energy range, with a slight departure from the experimental data at neutron energies below few meV. In particular, the AFGA seems to provide a much better agreement than the results based on direct DFT simulations~\cite{2019_Ramic_ANE} in the broad region below ca. 100 meV. To explain this result, one should consider that lucite is, in reality, a non-crystalline vitreous substance and, therefore, its simulation using DFT-based phonon calculations can be quite challenging, unless very large unit cells are considered. On the other hand, the AFGA provides the scattering contribution of a given functional group averaged over a series of slightly different environments and, therefore, is more able to reproduce the disordered nature of the material at the atomic scale.

The successful application of the AFGA to materials not related to amino acids of proteins is of particular interest for possible applications to a broader family of systems, including glassy materials, and prove its general applicability. In fact, while it is not expected for the AFGA to provide a high level of accuracy for small crystalline systems, the averaging of contributions of individual functional groups in a set of ``training'' systems can be representative of both large molecule, polymers and disordered biophysical systems. 

%-----------------
\subsection{Modelling using a sigmoidal function}
The definition of an average total cross section for hydrogen in organic systems, over the same energy range investigated here, was recently provided in Ref.~\cite{2019_Capelli_JAC}. In that case, the hydrogen total cross section was averaged over several organic molecules, including $\beta$-alanine, urea, and tartaric acid, so as to take into account the average contributions over different functional groups. In order to provide a simple analytical model, the result was fitted with a logistic 4-parameter sigmoidal function of the form
\begin{equation}
\sigma (E) \cong s_f + \frac{s_b}{1+cE^d} \quad .
\end{equation}
Considering the additional insight that we have gathered in this study, provided by the results of our {\it ab initio} calculations and the possibility to separate the signal from different functional groups, we have applied the same analytic function to each of 9 spectra obtained within the AFGA.
The parameters obtained for all types of functional groups are shown in the Table~\ref{Tab:Sigmoidal_parameters}. They allow a better treatment of neutron scattering experiments, providing detailed sample self-attenuation corrections for a variety of biological and soft-matter systems.
\begin{table}[htbp]
	\begin{center}
		\begin{tabular}{c|cccc} 
			\hline
  & s$_f$ & s$_b$ & c & d \\
\hline
\hline
CH(ali) & 19.14 $\pm$ 0.6 &  73.11  $\pm$ 1 &  37.46 $\pm$ 6 & 1.03 $\pm$ 0.04 \\
\hline
CH(aro) & 17.51 $\pm$ 0.7 &  86.72  $\pm$ 2 &  28.62 $\pm$ 4 & 0.84 $\pm$ 0.04 \\
\hline
CH$_2$ & 16.20 $\pm$ 0.9 &  105.1  $\pm$ 3 &  23.40 $\pm$ 2 & 0.71 $\pm$ 0.03 \\		
\hline
CH$_3$ & 13.60 $\pm$ 0.6 &  174.8  $\pm$ 7 &  27.14 $\pm$ 1 & 0.55 $\pm$ 0.02 \\
\hline
NH$_3$ & 18.82 $\pm$ 0.5 &  74.92  $\pm$ 1 &  32.90 $\pm$ 3 & 0.94 $\pm$ 0.03 \\
\hline
NH$_2$ & 17.54 $\pm$ 0.5 &  101.5  $\pm$ 2 &  28.75 $\pm$ 2 & 0.75 $\pm$ 0.02 \\
\hline	
NH & 19.10 $\pm$ 0.6 &  78.09  $\pm$ 1 &  36.39 $\pm$ 4 & 0.95 $\pm$ 0.03 \\	
\hline
OH & 19.27 $\pm$ 0.6 &  71.29  $\pm$ 1 &  44.57 $\pm$ 7 & 1.09 $\pm$ 0.05 \\
\hline
SH & 19.24 $\pm$ 0.4 &  107.3 $\pm$ 2 &  60.64 $\pm$ 5 & 0.85 $\pm$ 0.02 \\
			\hline
		\end{tabular}
	\end{center}
\caption{\label{Tab:Sigmoidal_parameters}Fitted curve parameters of different hydrogen contributions to the cross-section averaged over all amino acids using a sigmoidal function. }
\end{table}

\section{Conclusions}
We have provided experimental results for the total neutron cross section of the twenty proteinogenic amino acids in the neutron energy range between 1 meV and 10 keV. These were successfully reproduced applying the formalism of the multi-phonon expansion to the calculated vibrational densities of states obtained from {\it ab initio} simulations. Moreover, from the results of such calculations, we have defined the average contribution to the total cross section from hydrogen atoms in different functional groups. The set of 9 functions, defined by the average hydrogen dynamics in different chemical environments, was found to accurately reproduce the total cross section of all amino acids, in the framework named Average Functional Group Approximation, {\it i.e.,} AFGA. Moreover, we found that the AFGA could be successfully applied to triphenylmethane and poly-methylmethacrylate, or lucite, a glassy material for which phonon calculations are known to have limited success. 

Our results represent a considerable simplification in the challenging task of reproducing the total neutron cross section of dipeptides and proteins, with applications to biophysics and medicine. For example, our results can be utilised to provide new models for the macroscopic cross section of human tissues and muscles, and for applications to the cancer treatment through boron neutron capture therapy.

\section{Supplementary Material}
See the supplementary material for the values of the average hydrogen total cross section at 300 K and average VDoSs for functional groups in the AFGA model.

\section{Data Availability}
Raw data related to this article were generated at the ISIS Neutron and Muon Source (UK), DOI: 10.5286/ISIS.E.RB2010019. Derived data supporting the findings of this study are available from the corresponding author upon reasonable request.

\begin{acknowledgments}
The authors gratefully acknowledge the financial support of Regione Lazio (IR approved by Giunta Regionale n. G10795, 7 August 2019 published by BURL n. 69 27 August 2019), ISIS@MACH (I), and ISIS Neutron and Muon Source (UK) of Science and Technology Facilities Council (STFC); 
the financial support of Consiglio Nazionale delle Ricerche within CNR-STFC Agreement 2014-2020 (N 3420), concerning collaboration in scientific research at the ISIS Neutron and Muon Source (UK) of Science and Technology Facilities Council (STFC) is gratefully acknowledged.

The following article has been submitted to Journal of Chemical Physics. After it is
published, it will be found at https://aip.scitation.org/journal/jcp.

\end{acknowledgments}

\section*{References}

%\bibliography{biblio_Oct20}
%\include{NoCites}

%merlin.mbs aipnum4-1.bst 2010-07-25 4.21a (PWD, AO, DPC) hacked
%Control: key (0)
%Control: author (8) initials jnrlst
%Control: editor formatted (1) identically to author
%Control: production of article title (0) allowed
%Control: page (1) range
%Control: year (1) truncated
%Control: production of eprint (0) enabled
%

\end{document}